\documentclass[letterpaper,journal]{IEEEtran}

\usepackage{amsmath,amsfonts,amssymb}
\usepackage{algorithmic}
\usepackage{algorithm}
\usepackage{array}
\usepackage[caption=false,font=normalsize,labelfont=sf,textfont=sf]{subfig}
\usepackage{textcomp}
\usepackage{stfloats}
\usepackage{url}
\usepackage{verbatim}
\usepackage{graphicx}
\usepackage{cite}
\usepackage{multirow}
\usepackage{booktabs}
\usepackage{xcolor}
\usepackage{hyperref}
\usepackage{comment}
\usepackage{siunitx}

\hypersetup{
  pdftitle={TokAN: Accent Normalization Using Self-Supervised Speech Tokens},
  pdfauthor={Qibing Bai, Shuai Wang, Yuhan Du, Bohan Li, Yannan Wang, Haizhou Li},
  pdfsubject={IEEE Transactions on Audio, Speech, and Language Processing},
  pdfkeywords={accent conversion, discrete speech tokens, vector quantization, duration control, reinforcement learning}
}

\begin{document}

% ============================================================
%  TITLE AND AUTHORS
% ============================================================
\title{TokAN: Accent Normalization Using Self-Supervised Speech Tokens}

\author{Qibing~Bai,~\IEEEmembership{Student Member,~IEEE,}
        Shuai~Wang,~\IEEEmembership{Senior Member,~IEEE,}
        Yuhan~Du,
        Bohan Li,\\
        Yannan~Wang,
        and~Haizhou~Li,~\IEEEmembership{Fellow,~IEEE}%
\thanks{\emph{(Corresponding authors: Shuai Wang and Haizhou Li.)}}%
\thanks{Q.~Bai is with the School of Data Science (SDS), The Chinese University of Hong Kong, Shenzhen (CUHKSZ), China,
and with Tencent Ethereal Audio Lab, Tencent, Shenzhen, China.}%
\thanks{S.~Wang is with the School of Intelligence Science and Technology, Nanjing University, Suzhou, China, and with Shenzhen Loop Area Institute, Shenzhen, China.}%
\thanks{Y.~Du is with the School of Intelligence Science and Technology, Nanjing University, Suzhou, China.}%
\thanks{B.~Li is with the X-LANCE Lab, School of Computer Science, Shanghai Jiao Tong University, Shanghai, China.}%
\thanks{H.~Li is with School of Artificial Intelligence (SAI), CUHKSZ, China, with Shenzhen Research Institute of Big Data, China, and with Shenzhen Loop Area Institute, China.}%
\thanks{Y.~Wang is with Tencent Ethereal Audio Lab, Tencent, China.}%
\thanks{Samples: \url{https://p1ping.github.io/TokAN-Samples}.}
\thanks{Code: \url{https://github.com/P1ping/TokAN}.}}

% \markboth{IEEE Transactions on Audio, Speech, and Language Processing}%
\markboth{}%
{Bai \MakeLowercase{\textit{et al.}}: TokAN: Accent Normalization Using Self-Supervised Speech Tokens}

\maketitle

% ============================================================
%  ABSTRACT
% ============================================================
\begin{abstract}
Accent normalization (AN) seeks to convert non-native (L2) accented speech into standard (L1) speech while preserving speaker identity. The current techniques either require naturally recorded parallel L1--L2 speech for training, or suffer from quality degradation when supervised by synthesized targets.
In this paper, we present \textbf{TokAN}, a token-based accent normalization framework that operates on self-supervised discrete speech tokens extracted from a L1--L2 jointly trained vector-quantization (VQ) tokenizer, without the need of synthetic supervisory speech.
An autoregressive encoder-decoder model performs token-to-token conversion, translating L2-accented token sequences into the tokens of standard voice. We also introduce reinforcement learning (RL) post-training based on Group Relative Policy Optimization (GRPO), using word error rate and accent classifier confidence as complementary rewards.
A non-autoregressive flow-matching synthesizer recovers the Mel-spectrogram from the converted tokens, conditioned on the source speaker embedding.
We also develop a flow-matching duration predictor that supports total-duration-aware synthesis, making TokAN applicable to duration-critical tasks such as voice dubbing and live casting.
Experiments on seven English accents demonstrate that TokAN reduced the word error rate from 12.40\% to 9.89\% after supervised fine-tuning, and further to 9.23\% after RL post-training, consistently outperforming frame-to-frame, direct flow-matching, and prompt-based token-conversion baselines in terms of accent reduction and intelligibility.
\end{abstract}


% ============================================================
%  I. INTRODUCTION
% ============================================================
\section{Introduction}
\IEEEPARstart{A}{ccent} conversion (AC) seeks to alter speech from one accent to another while preserving the speaker's characteristics.
A particularly important special case is \emph{accent normalization} (AN), also referred to as foreign accent conversion (FAC)~\cite{zhao2019foreign}, which converts non-native (L2) accented speech into a native (L1) accented form.
AN technology enables a wide range of applications, including pronunciation training for language learners~\cite{felps2009foreign}, authentic multimedia dubbing~\cite{turk2002subband}, and personalized text-to-speech systems~\cite{sun2016personalized}.

Early deep learning approaches for AN are \emph{reference-based}~\cite{zhao2018icassp,zhao2019foreign,li2020improving,ding2022accentron}, relying on native accent speech samples to generate accent-neutral representations via phonetic posteriorgram (PPG) features~\cite{zhao2018icassp,zhao2019foreign,ding2022accentron} or native text-to-speech (TTS)~\cite{li2020improving}.
VEVO~\cite{zhang2025vevo} leverages speech tokens and accent prompts for prompt-based accent style transfer.
However, reference or prompt speech requirements at inference limit deployment in practice.

\emph{Reference-free} methods~\cite{zhao2021converting,nguyen2022accent,quamer2022zero} eliminate this requirement by directly mapping between accented and native accent representations~\cite{zhao2021converting,nguyen2022accent} or by converting PPGs for unseen speakers~\cite{quamer2022zero}.
While these approaches remove the inference-time dependency on reference speech, they still rely on naturally paired L1--L2 utterances or other paired supervision during training.
Subsequent work avoids the need for naturally recorded parallel corpora by constructing weakly paired or semi-synthetic supervision through cascaded ASR--TTS pipelines~\cite{liu2020end}, pseudo-Siamese network architectures with accent disentanglement~\cite{jin2023voice}, or TTS-guided accent-neutral representations~\cite{zhou2023tts,chen2024transfer}, further augmented with flow matching~\cite{bai2024diffusion} and normalizing flows~\cite{nguyen2024improving}.

Despite progress, methods relying on TTS-synthesized targets remain susceptible to two forms of quality degradation.
First, direct speech supervision propagates voice cloning artifacts and prosodic errors from synthetic targets into model training.
Second, duration information in synthetic speech may not reflect authentic native rhythm, injecting systematic errors into duration modeling.
\cite{jia2024convert}~addresses the parallel data requirement using discrete self-supervised tokens with a small amount of paired data, but does not eliminate the reliance on parallel corpora entirely.

Self-supervised discrete speech tokens~\cite{hsu2021hubert,chen2022wavlm} have emerged as a powerful intermediate representation that largely factors out speaker identity and duration while retaining phonetic content~\cite{choi24self}.
This property makes them attractive for accent normalization: operating at the token level naturally mitigates the impact of synthetic supervision quality, since token sequences encode far less paralinguistic information than raw waveforms or fine-grained codec representations~\cite{defossezhigh,du2024cosyvoice}.
Following this line, \cite{nguyen24syndata4genai}~explores synthetic data strategies for generalizable token-based conversion.
\cite{bai2025accent}~trains an autoregressive token conversion model on semi-synthetic token pairs constructed from non-parallel recordings, reducing the system's sensitivity to TTS-generated targets.

Despite the promise of token-based methods, two important challenges remain.
First, the discrete tokenizer itself plays a critical role: a tokenizer trained independently via K-Means clustering may not yield code assignments that are optimally discriminative for phonetic content relevant to accent conversion.
End-to-end joint training with the downstream synthesis objective can improve phonetic specificity and reconstruction fidelity.
Second, supervised fine-tuning (SFT) with token-level cross-entropy loss is a weak signal for the ultimate objectives of accent normalization---content preservation and accentedness reduction.
Reinforcement learning (RL) with task-relevant rewards can directly optimize these objectives and close the gap between SFT behavior and deployment criteria.

This paper presents \textbf{TokAN} (\textbf{Tok}en-based \textbf{A}ccent \textbf{N}ormalization), a comprehensive self-supervised-token-based accent normalization framework.
Compared with the preliminary conference version~\cite{bai2025accent}, which used HuBERT K-Means tokens, an accent-conditioned cross-attention converter, and supervised fine-tuning only, this journal version introduces three major technical extensions:
(i)~a jointly trained vector-quantization (VQ) tokenizer that co-optimizes phonetic content representation with speech synthesis and ASR-based supervision;
(ii)~a redesigned encoder-decoder conversion model with rotary positional encodings (RoPE), a self-attention-only decoder, and no source-accent embedding;
and (iii)~an RL post-training stage using GRPO~\cite{shao2024deepseek} with complementary WER-based and accent-classifier-based rewards.
In addition, the training pipeline is expanded in both data scale and diversity.
The non-autoregressive flow-matching synthesizer and total-duration-aware duration predictor are retained and analyzed in greater detail, completing the full system.

We evaluate TokAN on seven English accents from the L2-ARCTIC dataset~\cite{zhao2018l2arctic}, comparing against a frame-to-frame baseline (FramAN), a direct flow-matching method (CosyAccent~\cite{bai2026cosyaccent}), and a speech token-based system (VEVO~\cite{zhang2025vevo}).
The contributions of this paper are summarized as follows:
\begin{enumerate}
\item We replace the conference-version HuBERT/K-Means tokenizer with a jointly trained VQ tokenizer coupled to a speech synthesizer and a CTC-based ASR module, and conduct a systematic tokenizer selection study over SSL backbone, layer choice, and codebook size.
\item We redesign the autoregressive conversion model by adding RoPE, replacing cross-attention with a self-attention-only decoder that treats encoder features as a prefix, and removing the source accent embedding in favor of an accent-universal architecture.
\item We introduce an RL post-training stage based on GRPO that directly optimizes content preservation and accentedness reduction, using WER and accent classifier confidence as complementary rewards without requiring additional paired data.
\item We extend the duration-control study with a flow-matching duration predictor conditioned on the average token duration, providing a more detailed analysis of source-length preservation for dubbing and other duration-sensitive applications.
\item We substantially expand the training and evaluation beyond the conference version by using larger and more diverse datasets, adding stronger baselines, and incorporating more objective metrics.
\end{enumerate}

A preliminary version of this work was presented at Interspeech 2025~\cite{bai2025accent}.
The current paper is therefore a substantially revised system and evaluation centered on improved tokenization, accent-universal modeling, task-level post-training, larger-scale training, and broader empirical validation.

% ============================================================
%  II. RELATED WORK
% ============================================================
\section{Related Work}

\subsection{Accent Normalization}

Early deep-learning AN methods can be categorized by their dependence on reference speech and parallel data.
\emph{Reference-based} approaches use native speech samples at inference, generating accent-neutral representations via PPG or bottleneck features~\cite{zhao2018icassp,zhao2019foreign,ding2022accentron}, native TTS-generated references~\cite{li2020improving}, or speech tokens for style transfer~\cite{zhang2025vevo}.
\emph{Reference-free} methods eliminate this requirement by directly mapping between accented and native representations~\cite{zhao2021converting,nguyen2022accent}, by zero-shot PPG conversion~\cite{quamer2022zero}, or by exploiting discrete SSL tokens with limited paired data~\cite{jia2024convert}; see~\cite{huang2023evaluating} for a comprehensive evaluation.
The most data-flexible line drops real parallel L1--L2 pairs by constructing pseudo- or semi-synthetic supervision via cascaded ASR--TTS pipelines~\cite{liu2020end}, pseudo-Siamese architectures~\cite{jin2023voice}, or TTS-guided accent-neutral features~\cite{zhou2023tts,chen2024transfer}, augmented with flow matching~\cite{bai2024diffusion} or normalizing flows~\cite{nguyen2024improving}. CosyAccent~\cite{bai2026cosyaccent} introduces a direct flow-matching framework with a source-synthesis data strategy and explicit duration-ratio control, while FAC-FACodec~\cite{halychanskyi2025fac} explores accent-strength control via diffusion-timestep manipulation in a small-scale setup.
The current work falls into this semi-synthetic-pairing category for SFT, while its RL post-training stage further uses unpaired real multi-accented speech.

\subsection{Self-Supervised Discrete Speech Tokens}

Discrete tokens derived from SSL representations~\cite{hsu2021hubert,chen2022wavlm} correlate strongly with phonetic content~\cite{choi24self}, and have been adopted in voice conversion~\cite{huang2021any}, high-fidelity TTS~\cite{kharitonov2023speak}, spoken language modeling~\cite{lakhotia2021generative}, speech-to-speech translation~\cite{lee2022direct}, and LLM speech interfaces~\cite{fang2024llamaomni}.
While early systems use offline K-Means clustering on upper-layer features of HuBERT, WavLM, or W2V-BERT~\cite{lakhotia2021generative,kharitonov2023speak,chung2021w2v}, recent approaches adopt learnable codebooks jointly optimized with downstream objectives~\cite{du2024cosyvoice,du2024cosyvoice2}.
Our tokenizer follows this direction, co-training the codebook with synthesis and ASR objectives.

\subsection{Flow Matching and RL in Speech Generation}

Flow matching~\cite{lipman2023flow} enables straighter probability paths and more efficient sampling than diffusion, achieving competitive naturalness in TTS~\cite{mehta2024matcha} and supporting flow-matching duration predictors~\cite{le2023voicebox,ju2024naturalspeech,eskimez24total}.
We apply flow matching to both the token-to-Mel synthesizer and the duration predictor.
RL-based post-training has improved LLMs~\cite{ouyang2022training,shao2024deepseek} and recently been applied in speech to ASR~\cite{shivakumar2025grpoasr} and TTS prosody~\cite{liu2021reinforcement,li2026dmospeech2}.
GRPO~\cite{shao2024deepseek} extends PPO~\cite{schulman2017proximal} by eliminating the value network in favor of group-relative advantages, making it well-suited for sequence generation.
To our knowledge, this is the first application of GRPO post-training to accent conversion.

% ============================================================
%  III. SYSTEM OVERVIEW
% ============================================================
\section{System Overview}

\begin{figure*}[t]
\centering
\includegraphics[width=0.9\textwidth]{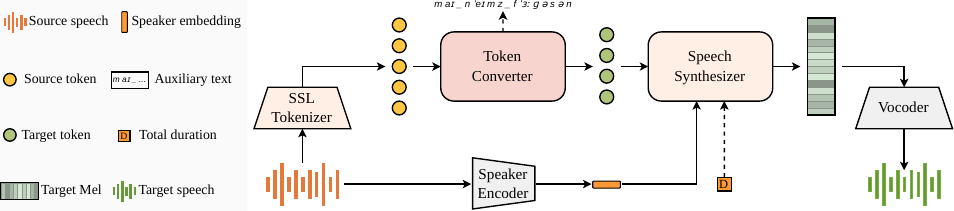}
\caption{Overview of TokAN. (a)~The SSL tokenizer encodes L2-accented speech into discrete tokens via a jointly trained VQ codebook. (b)~The autoregressive token converter transforms L2-accented tokens into L1-accented ones. (c)~The speech synthesizer generates the Mel-spectrogram from the converted tokens, conditioned on the source speaker embedding and an optional total-duration constraint. A pre-trained vocoder synthesizes the final waveform.}
\label{fig:pipeline}
\end{figure*}

Figure~\ref{fig:pipeline} illustrates the complete TokAN pipeline, which consists of three independently trainable components.

\textbf{VQ Tokenizer.}
A frozen WavLM-Large model~\cite{chen2022wavlm} extracts frame-level representations from the input waveform.
A learnable VQ codebook maps these continuous representations to discrete tokens.
The tokenizer is jointly trained with a flow-matching speech synthesizer and a CTC-based ASR module on native English speech, ensuring that the codebook captures phonetically rich representations suitable for both content preservation and accurate reconstruction.

\textbf{Token Conversion Model.}
An autoregressive encoder-decoder model transforms the source (L2-accented) token sequence into a target (native-accented) token sequence.
The encoder is a Transformer~\cite{vaswani2017attention} with rotary positional encodings (RoPE)~\cite{su2024roformer}, supervised by CTC phone prediction to reinforce phonemic content understanding.
The decoder is self-attention-only, with the encoder output prepended as a prefix in the sequence dimension, eliminating separate cross-attention layers.
The conversion model is first pre-trained with BART-style corruption on unlabeled native speech, then fine-tuned on semi-synthetic parallel token pairs.
An optional RL post-training stage using GRPO further refines the decoder toward content preservation and accent reduction.

\textbf{Flow-Matching Synthesizer.}
A non-autoregressive synthesizer maps the converted native token sequence to a Mel-spectrogram, conditioned on a speaker embedding extracted from the source speech.
A flow-matching duration predictor assigns durations to each token; when conditioned on the average token duration, it supports total-duration-aware generation.
The Mel-spectrogram is converted to a waveform using a pre-trained HiFT vocoder~\cite{li2023hiftnet}.

Crucially, since the VQ tokenizer serves as the interface between the synthesizer and the conversion model, the two modules can be trained on entirely different datasets, enabling modular optimization: the synthesizer is trained on high-quality native speech only, while the conversion model is trained on semi-synthetic token pairs constructed from non-parallel corpora.

% ============================================================
%  IV. VQ TOKENIZER
% ============================================================
\section{Joint Quantization \& Synthesis}

\subsection{Motivation}

Prior work on discrete-token speech conversion~\cite{bai2025accent,jia2024convert} extracts SSL features and applies offline K-Means clustering to obtain discrete tokens.
This approach has two limitations.
First, K-Means is trained independently of the downstream tasks (synthesis and conversion), so the cluster assignments are not directly optimized for phonetic discriminability or reconstruction quality.
Second, the codebook is static after clustering, precluding end-to-end gradient flow.

We replace K-Means with a learnable vector quantization (VQ) codebook~\cite{van2017vq}, jointly optimized with a flow-matching speech synthesizer and a CTC-based ASR module.
This joint training encourages the codebook to produce token sequences that are simultaneously easy to reconstruct into high-quality speech. Ideally, the reconstruction/resynthesis should achieve lower WERs for both L1 and L2 accents.

\begin{figure}[ht]
\centering
\includegraphics[width=0.485\textwidth]{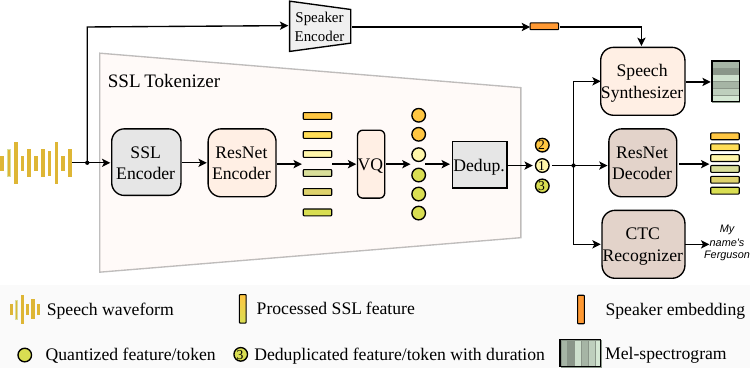}
\caption{Architecture of the jointly trained SSL tokenizer and speech synthesizer. A frozen SSL encoder extracts continuous features, which pass through a shallow 1-D ResNet and VQ bottleneck to produce discrete tokens. After deduplication, the token embeddings (with their durations) are fed to three modules: (a)~a flow-matching speech synthesizer that generates Mel-spectrograms conditioned on speaker embedding and optional total-duration input, (b)~a ResNet decoder for SSL feature reconstruction, and (c)~a CTC recognizer for content supervision. During joint training, tokens are represented as their corresponding codebook vectors to enable gradient flow.}
\label{fig:tokenizer}
\end{figure}

\subsection{Overall Architecture and Joint Training}
\label{sec:joint_tokenizer_arch}

As shown in Figure~\ref{fig:tokenizer}, the tokenizer comprises four modules: (i)~a frozen SSL encoder providing frame-level features at a deep layer, where strong phonetic encoding has been observed~\cite{choi24self}; (ii)~a shallow 1-D ResNet encoder--decoder pair surrounding a \textbf{VQ bottleneck} with codebook size $V$; (iii)~the flow-matching speech synthesizer detailed in Sec.~\ref{sec:synthesizer}; and (iv)~a Transformer-based CTC character recognizer.
Within the VQ bottleneck, each frame-level feature $\mathbf{f}$ is mapped to its nearest codebook entry:
\begin{equation}
k^* = \arg\min_{k \in \{1,\ldots,V\}} \| \mathbf{f} - \mathbf{e}_k \|_2, \quad \mathbf{z}_q = \mathbf{e}_{k^*}
\end{equation}
where $\{\mathbf{e}_k\}_{k=1}^V$ is the codebook of size $V$.
Straight-through estimation~\cite{van2017vq} passes gradients through the non-differentiable argmin, and the VQ-VAE objective is
\begin{equation}
  \mathcal{L}_\text{VQ} = \mathcal{L}_\text{recon} + \beta \|\mathbf{f} - \text{sg}(\mathbf{z}_q)\|_2^2
\label{eq:vq_loss}
\end{equation}
where $\mathcal{L}_\text{recon}$ is the L1 reconstruction loss between the ResNet decoder output and the original SSL features (i.e., the input to the 1-D ResNet encoder, before any processing), and the second term is the commitment loss with stop-gradient $\text{sg}(\cdot)$. The codebook is updated via EMA rather than gradient descent (see~\cite{van2017vq}), with $\beta = 0.5$ and EMA decay $\alpha = 0.99$.
The synthesizer objective $\mathcal{L}_\text{synth} = \mathcal{L}_\text{speech} + \mathcal{L}_\text{duration}$ aggregates the flow-matching losses for the Mel-spectrogram and token-level durations, while the recognizer contributes a CTC loss $\mathcal{L}_\text{recog}$.

The joint training objective combines three losses:
\begin{equation}
\mathcal{L}_{\text{tokenizer}} = \lambda_1 \mathcal{L}_{\text{VQ}} + \lambda_2 \mathcal{L}_{\text{synth}} + \lambda_3 \mathcal{L}_{\text{recog}}
\label{eq:tok_loss}
\end{equation}
with $\lambda_1 = 2.0$, $\lambda_2 = 2.0$, and $\lambda_3 = 0.5$.

After quantization, consecutive identical tokens are merged (deduplication) before being passed to the conversion model. This removes duration information so that conversion focuses on phonetic mapping, and enables token-level duration manipulation in the synthesis stage.

\subsection{Synthesizer Architecture and Inference}
\label{sec:synthesizer}

The speech synthesizer is a non-autoregressive model that maps the converted (deduplicated) native token sequence $\hat{\mathbf{y}}$ to a Mel-spectrogram, conditioned on a speaker embedding $\mathbf{s}$ extracted from the source speech using Resemblyzer\footnote{\url{https://github.com/resemble-ai/Resemblyzer}}.
The synthesizer consists of three components: (i)~a Transformer-based token encoder, (ii)~a flow-matching duration predictor, and (iii)~a DiT-based Mel decoder.

\textbf{Token encoder.}
A Transformer with relative positional embeddings maps the deduplicated token embedding sequence to continuous representations, conditioned on speaker embeddings via AdaLN~\cite{peebles2023scalable}.

\textbf{Mel decoder.}
The duration predictor assigns integer durations to each token; the token representations are expanded accordingly and serve as condition for the Mel decoder.
The Mel decoder follows the diffusion Transformer (DiT) architecture~\cite{peebles2023scalable} and transforms a noisy Mel-spectrogram into a high-likelihood one via velocity prediction.
Speaker conditioning uses AdaLN-Zero~\cite{peebles2023scalable,perez2018film}.
Classifier-free guidance (CFG)~\cite{ho2021classifierfree} is applied for both content and speaker conditions:
\begin{align}
\bar{v}_\eta(\mathbf{m}_t, t, \hat{\mathbf{y}}, \mathbf{s}) &= v_{\eta}(\mathbf{m}_t, t, \hat{\mathbf{y}}, \mathbf{s}) \nonumber \\
&+ w_1\left[v_{\eta}(\mathbf{m}_t, t, \hat{\mathbf{y}}, \mathbf{s}) - v_{\eta}(\mathbf{m}_t, t, \varnothing, \mathbf{s})\right] \nonumber \\
&+ w_2\left[v_{\eta}(\mathbf{m}_t, t, \hat{\mathbf{y}}, \mathbf{s}) - v_{\eta}(\mathbf{m}_t, t, \hat{\mathbf{y}}, \varnothing)\right]
\label{eq:cfg}
\end{align}
where $\mathbf{m}_t$ is the noisy Mel-spectrogram state at flow time $t$, $\hat{\mathbf{y}}$ is converted tokens after encoding, and $w_1$ and $w_2$ control the emphasis on the content and timbre conditions, respectively.
A vocoder then synthesizes the waveform $\hat{\mathbf{o}}$ given $\mathbf{m}_1$.

\textbf{Duration prediction and total duration control.}
\label{sec:duration}
The duration predictor is a flow-matching model conditioned on the token representations and speaker embedding.
By default (duration-free mode), it predicts token-level durations without a total-duration constraint.

For applications requiring total-duration preservation (e.g., dubbing), we condition the duration predictor on the \textit{average token duration} $\bar{d} = D_\text{tgt} / L_\text{tgt}$, where $D_\text{tgt}$ is the target total duration and $L_\text{tgt}$ is the target sequence length.
This provides direct soft control over the output duration while allowing the predictor to allocate durations flexibly across tokens.
The total-duration condition is randomly dropped in training so that the duration predictor supports both the duration-free and source-length modes.
For the source-length mode, a CFG rate $w_3 = 0.1$ is deployed to emphasize total-duration awareness.

We denote TokAN in the two modes as:
\begin{itemize}
\item \textbf{TokAN-1} (duration-free): duration predictor without total-duration conditioning.
\item \textbf{TokAN-2} (source-length): duration predictor conditioned on the source average token duration.
\end{itemize}

\subsection{Tokenizer Selection Study}
\label{sec:tokenizer_selection}

We conduct a systematic study to identify the optimal tokenizer configuration.
The selection criterion is the word error rate (WER) computed by a native-only ASR model on speech reconstructed from discrete tokens---a proxy for the tokenizer's joint content preservation and reconstruction capability.
Additionally, we consider the initial accent normalization ability, i.e., the extent to which the reconstructed speech has reduced accentedness even before any conversion, which we estimate by the same native-only WER metric on L2-accented speech reconstructed through the tokenizer. We focus on the accent-removing speech reconstruction capability and, therefore, use the source token durations instead of predicting them.

\subsubsection{SSL Model and Layer}

We compare three SSL backbones at two commonly used layers:
\begin{itemize}
\item \textbf{Wav2vec 2.0}\footnote{\url{https://huggingface.co/facebook/wav2vec2-large-lv60}}~\cite{baevski2020wav2vec}: layers 17 and 22.
\item \textbf{HuBERT-large}\footnote{\url{https://huggingface.co/facebook/hubert-large-ll60k}}~\cite{hsu2021hubert}: layers 17 (used in the preliminary conference version~\cite{bai2025accent}) and 22.
\item \textbf{WavLM-Large}\footnote{\url{https://huggingface.co/microsoft/wavlm-large}}~\cite{chen2022wavlm}: layers 17 and 22.
\item \textbf{W2V-BERT 2.0}\footnote{\url{https://huggingface.co/facebook/w2v-bert-2.0}}~\cite{chung2021w2v}: layers 17 and 22.
\end{itemize}
All configurations use a codebook of size 1024 and are trained on LibriTTS-R~\cite{koizumi2023librittsr}.
Table~\ref{tab:tokenizer_model} reports the reconstruction WER on the L2-ARCTIC validation set. We adopt a native-only ASR model\footnote{\url{https://huggingface.co/facebook/s2t-medium-librispeech-asr}}, simulating listener's perception of L2-accented speech. This can be considered an intelligibility metric reflecting both content preservation and nativeness.

\begin{table}[!t]
\caption{Tokenizer Selection: SSL Model and Layer (Codebook Size = 1024, Trained on LibriTTS-R). WER (\%) computed by native-only ASR.}
\label{tab:tokenizer_model}
\centering
\small
\setlength{\tabcolsep}{5pt}
\begin{tabular}{lccc}
\toprule
\multirow{2}{*}{SSL Model} & \multirow{2}{*}{Layer} & \multicolumn{2}{c}{WER (\%) $\downarrow$} \\
\cmidrule(r){3-4}
& & Overall & L2 Accents \\
\midrule
\multirow{1}{*}{Source} & - & 19.81 & 22.50 \\
\midrule
\multirow{2}{*}{Wav2vec 2.0} & 17 & 22.33 & 25.24 \\
                              & 22 & 87.88 & 90.08 \\
\midrule
\multirow{2}{*}{HuBERT} & 17 & 21.29 & 24.02 \\
                              & 22 & 18.97 & 21.46 \\
\midrule
\multirow{2}{*}{WavLM}  & 17 & 18.71 & 21.16 \\
                              & 22 & \bf{17.45} & \bf{19.82} \\
\midrule
\multirow{2}{*}{W2V-BERT 2.0} & 17 & 21.55 & 24.42 \\
                              & 22 & 22.76 & 25.89 \\
\bottomrule
\end{tabular}
\end{table}

WavLM-Large at layer~22 achieves the lowest reconstruction WER, confirming its stronger phonetic encoding and accent robustness~\cite{chen2022wavlm}.
We fix the SSL backbone to WavLM-Large layer~22 for all subsequent experiments.

\begin{table}[!t]
\caption{Tokenizer Selection: Codebook Size (WavLM-Large, Layer~22). WER (\%) by native-only ASR.}
\label{tab:tokenizer_codebook}
\centering
\small
\setlength{\tabcolsep}{5pt}
\begin{tabular}{lcc}
\toprule
\multirow{2}{*}{Codebook Size} & \multicolumn{2}{c}{WER (\%) $\downarrow$} \\
\cmidrule(r){2-3}
& Overall & L2 Accents \\
\midrule
256  & 18.02 & 20.35 \\
512  & 17.95 & 20.28 \\
1024 & \bf{17.45} & \bf{19.82} \\
2048 & 17.64 & 19.95 \\
4096 & 18.27 & 20.53 \\
\bottomrule
\end{tabular}
\end{table}

\subsubsection{Codebook Size}

Fixing the SSL backbone to WavLM-Large layer~22, we vary the codebook size $V \in \{256, 512, 1024, 2048, 4096\}$.
Table~\ref{tab:tokenizer_codebook} reports the results.
A codebook of size 1024 achieves the best balance between expressiveness and normalization.
We therefore fix $V = 1024$ for the final model.

\subsubsection{Training Strategy / Ablation Study}

\begin{table}[!t]
\caption{Tokenizer Training Strategies.}
\label{tab:tokenizer_ablation}
\centering
\small
\setlength{\tabcolsep}{5pt}
\begin{tabular}{llcc}
\toprule
\multirow{2}{*}{Strategy} & \multirow{2}{*}{Objective} & \multicolumn{2}{c}{WER (\%) $\downarrow$} \\
\cmidrule(r){3-4}
& & Overall & L2 Accents \\
\midrule
\multirow{2}{*}{Joint Tok. \& Synth.} & $\mathcal{L}_\text{tokenizer}$ & \bf{17.45} & \bf{19.82} \\
                              & w/o $\mathcal{L}_\text{recog}$ & 17.90 & 20.13 \\
\midrule
\multirow{2}{*}{Separate Tok. \& Synth.} & VQ ($\mathcal{L}_\text{VQ}$) & 18.52 & 20.85 \\
                              & K-Means & 19.63 & 22.14 \\
\bottomrule
\end{tabular}
\end{table}

Our tokenizer is jointly optimized for 1) SSL feature quantization and reconstruction, 2) flow-matching speech synthesis, and 3) CTC-based speech recognition. To validate the effectiveness of this joint training strategy, we perform a simple ablation study. We firstly remove the auxiliary recognition objective (second row) and the WER increases, indicating the auxiliary CTC objective helps the joint quantization-synthesis pipeline.

We further test the separate training strategy -- simply training a quantization module and then training a synthesizer on the extracted tokens. As the third row shows, such a separate training strategy has inferior performance. We also test the K-Means method to train the quantization model, which we previously deployed in the conference version. As the 4th row shows, the VQ method has better performance, likely due to its larger number of parameters and longer training time, which enable better data fitting.

After completing the selection study, we retrain the joint VQ tokenizer and flow-matching synthesizer on the larger Emilia-EN dataset~\cite{he2024emilia} with the selected configuration (WavLM-Large, layer~22, $V=1024$) to maximize coverage of diverse native speech patterns.

% ============================================================
%  V. AUTOREGRESSIVE TOKEN CONVERSION
% ============================================================
\section{Autoregressive Token Conversion}

\begin{figure}[ht]
\centering
\includegraphics[width=0.46\textwidth]{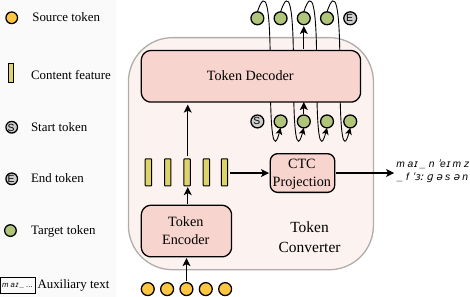}
\caption{Architecture of the token converter. The encoder consumes source speech tokens and produces content representations, supervised by a CTC phoneme loss. The encoder output is prepended as a prefix to the self-attention decoder, which autoregressively generates L1-accented speech tokens.}
\label{fig:converter}
\end{figure}

The token conversion model has an encoder-decoder architecture, as shown in Fig.~\ref{fig:converter}.
Both encoder and decoder are Transformers equipped with RoPE~\cite{su2024roformer}
, which provides relative positional awareness without absolute position embeddings, improving generalization to variable-length sequences.

\textbf{Encoder.}
The Transformer encoder maps the deduplicated L2-accented token sequence $\mathbf{x} = (x_1, \ldots, x_S)$ to a sequence of continuous content representations $\mathbf{c} = (c_1, \ldots, c_S)$:
\begin{equation}
\mathbf{c} = \mathrm{Encoder}(\mathrm{Embed}(\mathbf{x}))
\end{equation}

To encourage the encoder to produce phonemically informative representations, we attach a linear projection head to the encoder outputs and compute a CTC loss~\cite{graves2006connectionist} against phoneme label sequences derived from text transcripts:
\begin{equation}
\mathcal{L}_{\mathrm{CTC}} = \mathrm{CTC}(\mathrm{Linear}(\mathbf{c}),\, \mathbf{p})
\end{equation}
where $\mathbf{p}$ is the reference phoneme sequence.
This auxiliary task explicitly encourages the encoder to maintain content-related structure, aiding the recognition-generation pipeline.

\textbf{Decoder.}
Unlike the preliminary version~\cite{bai2025accent}, which uses separate cross-attention to condition on encoder outputs, the decoder here is \textbf{self-attention-only}.
The encoder content representations $\mathbf{c}$ are prepended to the target token embeddings as a prefix in the sequence dimension:
\begin{equation}
\mathbf{d} = \mathrm{Decoder}\left([\mathbf{c};\, \mathrm{Embed}(\mathbf{y}_{<t})]\right)
\end{equation}
where $[\cdot\,;\,\cdot]$ denotes sequence concatenation and $\mathbf{y}_{<t}$ is the target token history.
The decoder attends to the entire prefix (encoder representations) as well as the preceding target tokens via causal self-attention.
This design unifies the conditioning and generation paths into a single attention mechanism, which has been shown to improve efficiency and representational capacity in recent language models.

We remove the source accent embedding used in the conference version~\cite{bai2025accent}, making the model \textit{accent-universal}.
The accent information is instead encoded directly in the encoder's input sequence, and the model is expected to handle diverse L2 accents from the phonetic content alone.

\subsection{BART-style Pre-training}
\label{sec:pretraining}

To initialize the conversion model with priors over diverse token distributions, we apply BART-style~\cite{lewis2019bart,jia2024convert} pre-training on unlabeled speech from Emilia-EN~\cite{he2024emilia}.
This task simulates speech token generation from a non-standard sequence, similar to accent normalization.
Given a clean native token sequence, we iterate through positions and apply one of three operations: \textbf{span masking} (probability $p_\text{mask}$), where a Poisson-length span is either collapsed into a single \texttt{[MASK]} token or, with probability $p_\text{rand}$, replaced with random tokens drawn uniformly from the vocabulary; \textbf{insertion} (probability $p_\text{ins}$), where a random token is prepended to the current position; and \textbf{keep} otherwise.
Because span collapsing changes the input length, the model must jointly recover content and length, closely matching the source-vs-target length mismatch encountered in accent conversion.

\subsection{Supervised Fine-Tuning}
\label{sec:sft}

After pre-training, we fine-tune the conversion model on semi-synthetic parallel token pairs constructed from corpora that are not naturally parallel (see Sec.~\ref{sec:datasets}).
Both the L2-accented source and the native-accented target are tokenized using the VQ tokenizer.
The model is trained with a cross-entropy loss on the token sequence:
\begin{equation}
\mathcal{L}_{\text{token}} = -\sum_{t=1}^T \log p_\theta(y_t \mid \mathbf{x}, \mathbf{y}_{<t})
\end{equation}
combined with the CTC phoneme guidance loss:
\begin{equation}
\mathcal{L}_{\text{SFT}} = \mathcal{L}_{\text{token}} + \gamma \mathcal{L}_{\mathrm{CTC}}
\label{eq:sft_loss}
\end{equation}
with auxiliary CTC weight $\gamma = 0.5$.

Since the source utterances in the fine-tuning data are mostly L2-accented and the targets are TTS-synthesized, both the source and target token sequences are extracted from the VQ tokenizer rather than from raw speech.
Operating at the token level significantly reduces the impact of voice cloning artifacts in synthetic targets, as such artifacts primarily manifest in paralinguistic features (e.g., fine-grained spectral details) that are largely abstracted away by the SSL tokenizer.

% ============================================================
%  VI. RL POST-TRAINING
% ============================================================
\section{Reinforcement Learning Post-Training}

\subsection{Motivation}

Supervised fine-tuning optimizes a proxy objective (next-token prediction) that does not directly measure the quality of the complete output sequence in terms of accent normalization.
Two key discrepancies arise: (i)~WER (content preservation) depends on the entire generated sequence, not individual token decisions; and (ii)~accentedness reduction is not well captured by token-level cross-entropy.
RL post-training allows the model to directly optimize these task-level objectives through reward signals computed on complete output sequences.

Crucially, RL post-training with rewards based on ASR and accent classifier does not require any additional paired data: the model explores the output space by sampling, and rewards are computed fully automatically on the sampled outputs.

\subsection{GRPO Formulation}
\label{sec:grpo}

We apply Group Relative Policy Optimization (GRPO)~\cite{shao2024deepseek} to fine-tune the decoder.
The encoder and token embedding table are frozen to maintain the content-extraction behavior of the encoder; only the decoder parameters $\theta$ are updated.

For each source utterance $\mathbf{x}$, we sample a group of $G = 12$ output sequences from the current policy using nucleus (top-$p$) and top-$k$ sampling with $p = 0.85$ and $k = 25$:
\begin{equation}
\{\mathbf{y}^{(g)}\}_{g=1}^G \sim p_\theta(\cdot \mid \mathbf{x})
\end{equation}
Rewards $r^{(g)}$ are computed for each sample (Sec.~\ref{sec:rewards}).
Group-relative advantages are computed as:
\begin{equation}
A^{(g)} = \frac{r^{(g)} - \mathrm{mean}(\{r^{(g')}\}_{g'=1}^G)}{\mathrm{std}(\{r^{(g')}\}_{g'=1}^G) + \epsilon_A}
\end{equation}
The policy gradient objective with clipping (cf. PPO~\cite{schulman2017proximal}) is:
\begin{align}
\mathcal{L}_{\text{GRPO}}(\theta) = -\mathbb{E} \big[ \min \big( &\rho^{(g)} A^{(g)}, \nonumber \\
&\mathrm{clip}(\rho^{(g)}, 1-\epsilon, 1+\epsilon)\, A^{(g)} \big) \big]
\end{align}
where $\rho^{(g)} = p_\theta(\mathbf{y}^{(g)} \mid \mathbf{x}) / p_{\theta_\text{old}}(\mathbf{y}^{(g)} \mid \mathbf{x})$ is the importance weight and $\epsilon = 0.2$ is the clip range.
In practice, we adopt DAPO-style loss~\cite{yu2026dapo} implemented in TRL\footnote{\url{https://github.com/huggingface/trl}}, which averages the clipped objective uniformly across all tokens in the group rather than first averaging within each sample, assigning equal weight to every token regardless of sequence length.
A KL penalty term is also added to prevent the policy from deviating too far from the SFT policy:
\begin{equation}
\mathcal{L}_{\text{rl}}(\theta) = \mathcal{L}_{\text{GRPO}}(\theta) + \lambda_{\text{kl}} \, D_{\mathrm{KL}}\!\left(p_\theta(\cdot \mid \mathbf{x}) \,\|\, p_{\theta_{\mathrm{sft}}}(\cdot \mid \mathbf{x})\right)
\end{equation}
with the regularization weight $\lambda_{\text{kl}} = 0.02$.

During RL training, token sequences are generated in the total-duration-free mode, i.e., the conversion model predicts the target token sequence without a total-duration constraint.
The synthesizer and vocoder convert the sampled token sequences to waveforms for reward computation. For faster generation, the synthesizer uses 10 steps and a simpler CFG scheme:
$\bar{v}_\eta(\mathbf{m}_t, t, \hat{\mathbf{y}}, \mathbf{s}) = 2\, v_{\eta}(\mathbf{m}_t, t, \hat{\mathbf{y}}, \mathbf{s}) - v_{\eta}(\mathbf{m}_t, t, \varnothing, \varnothing)$.

\subsection{Reward Design}
\label{sec:rewards}

We design two complementary rewards, targeting content preservation and accent reduction, respectively.

\textbf{WER-based reward.}
Content preservation is measured using a native-only ASR model (Whisper-medium-EN~\cite{radford2023robust}):
\begin{equation}
r_{\mathrm{WER}}(\mathbf{y}) = 1 - \mathrm{WER}(\hat{\mathbf{w}}, \mathbf{w}^*)
\end{equation}
where $\hat{\mathbf{w}}$ are the ASR hypotheses from the synthesized output and $\mathbf{w}^*$ are the ground-truth transcripts.
A higher reward corresponds to better intelligibility.

\textbf{Accent-based reward.}
Accentedness reduction is measured using an accent classifier~\cite{juan2023commonaccent}\footnote{\url{https://huggingface.co/Jzuluaga/accent-id-commonaccent_xlsr-en-english}} pre-trained on Common Voice 7.0~\cite{ardila2020commonvoice}.
Among the classifier's 16 accent labels, we treat US and England as native targets, and define the reward as the sum of their predicted probabilities:
\begin{equation}
r_{\text{Acc}}(\mathbf{y}) = P(\mathrm{accent} \in \{\text{US}, \text{England}\} \mid \hat{\mathbf{o}})
\end{equation}
where $\hat{\mathbf{o}}$ is the synthesized speech waveform.
This reward is complementary to $r_{\mathrm{WER}}$: it directly captures accentedness without being confounded by transcription errors.

The combined reward is a weighted sum:
\begin{equation}
r(\mathbf{y}) = w_{\mathrm{WER}} \cdot r_{\mathrm{WER}}(\mathbf{y}) + w_{\text{Acc}} \cdot r_{\text{Acc}}(\mathbf{y})
\label{eq:reward}
\end{equation}
with reward weights $w_{\mathrm{WER}} = 1.0$ and $w_{\text{Acc}} = 0.5$.

\subsection{Training Data for RL}

We use the GLOBE multi-accent dataset~\cite{wang2024globe}\footnote{\url{https://huggingface.co/datasets/MushanW/GLOBE}} for RL post-training.
GLOBE provides real multi-accented speech utterances from diverse speakers and accent backgrounds, but no paired native targets are required---the model explores freely and receives rewards from the ASR and accent classifier modules.
This on-policy exploration on real multi-accented data ensures that the model improves on the actual distribution of inputs encountered at deployment, rather than on the semi-synthetic distribution used in SFT.

% ============================================================
%  VII. TRAINING PIPELINE
% ============================================================
\section{Training Pipeline}

The complete training pipeline consists of four stages:

\begin{enumerate}
\item \textbf{VQ tokenizer and synthesizer joint training.}
The VQ codebook, flow-matching synthesizer, and auxiliary CTC-based ASR module are jointly trained on LibriTTS-R~\cite{koizumi2023librittsr} for the selection study, and subsequently retrained on Emilia-EN~\cite{he2024emilia} with the selected configuration.
The SSL backbone (WavLM-Large) is frozen throughout.
The pre-trained 50\,Hz HiFT vocoder from CosyVoice2~\cite{du2024cosyvoice2} is used for waveform generation.

\item \textbf{BART-style pre-training of the conversion model.}
The encoder-decoder model is pre-trained on diverse English speech from Emilia-EN using the BART-style corruption objective (Sec.~\ref{sec:pretraining}).
This stage familiarizes the model with diverse token distributions and trains the encoder with CTC phone supervision.

\item \textbf{Supervised fine-tuning (SFT).}
The conversion model is fine-tuned on two semi-synthetic datasets:
(i)~L2-LibriTTSR~\cite{bai2026cosyaccent}: synthesized non-native utterances (L2 source) paired with their original LibriTTS-R recordings (native target);
(ii)~extended L2-ARCTIC~\cite{zhao2018l2arctic}: real L2-accented speech paired with TTS-synthesized native targets.
Both datasets provide source-target token pairs for cross-entropy fine-tuning (Eq.~\ref{eq:sft_loss}).

\item \textbf{GRPO post-training.}
The encoder and embedding table are frozen.
Only the decoder is fine-tuned using GRPO on real accent-diverse utterances from GLOBE~\cite{wang2024globe}, with the combined WER$+$accent reward (Eq.~\ref{eq:reward}).
\end{enumerate}

The four stages are performed sequentially.

% ============================================================
%  IX. EXPERIMENTAL SETUP
% ============================================================
\section{Experimental Setup}

\subsection{Datasets}
\label{sec:datasets}

\textbf{LibriTTS-R}~\cite{koizumi2023librittsr}: High-quality native English speech dominated by the US accent.
Used to select the configuration of the jointly trained SSL tokenizer and synthesizer.

\textbf{Emilia-EN}~\cite{he2024emilia}: Large-scale English speech dataset covering diverse speaking styles and accents.
Used for BART pre-training and jointly training the SSL tokenizer and synthesizer.

\textbf{L2-LibriTTSR}~\cite{bai2026cosyaccent}: Synthesized non-native counterparts of LibriTTS-R utterances, covering multiple L2 accents.
Used for SFT, together with LibriTTS-R.

\textbf{L2-ARCTIC}~\cite{zhao2018l2arctic} + \textbf{ARCTIC}~\cite{kominek2004cmu}: Six non-native English accents (Arabic, Chinese, Hindi, Korean, Spanish, Vietnamese) with four speakers each; native US English speakers from ARCTIC.
Used for SFT and evaluation.
Native targets for L2-ARCTIC are synthesized using a zero-shot Matcha-TTS~\cite{mehta2024matcha} trained on LibriTTS-R.
We use the same 50/80 sentence split for validation/testing, consistent with~\cite{bai2026cosyaccent}.

\textbf{GLOBE}~\cite{wang2024globe}: Multi-accent English speech corpus with diverse L1 and L2 speakers.
Used for GRPO post-training.
No paired native targets are required.

\subsection{Compared Systems}
\label{sec:systems}

We evaluate TokAN against three baselines.

\textbf{FramAN} (Frame-to-frame Accent Normalization)~\cite{bai2024diffusion}: A three-stage frame-to-frame system. (i)~A FastSpeech-like~\cite{ren2021fastspeech} non-autoregressive TTS model is trained on Emilia-EN, providing an accent-neutral text encoder and a flow-matching acoustic decoder; unlike~\cite{bai2024diffusion}, we omit intermediate pitch/energy variables, which empirically degrade speech quality. (ii)~A speech encoder distills Whisper-medium~\cite{radford2023robust} features to match the text encoder output, with alignment from bournemouth-forced-aligner~\cite{rehman2025bfa} and modality-matching loss on Emilia-EN. (iii)~The speech encoder is fine-tuned on phoneme-aligned native targets synthesized by the TTS branch, combining modality-matching and flow-matching losses.
FramAN always preserves the source rhythm, including the total duration.

\textbf{CosyAccent}~\cite{bai2026cosyaccent}: A direct non-autoregressive flow-matching model with a duration-ratio predictor. Before finetuning on the semi-synthesized data, we pretrain it with a stop-gradient operation on its encoder outputs. We evaluate two modes of CosyAccent: \textit{CosyAccent-1} (duration-free) and \textit{CosyAccent-2} (source-length). 

\textbf{VEVO}~\cite{zhang2025vevo}: An autoregressive token-level conversion model that uses codebook size as an information bottleneck---a small-codebook HuBERT tokenizer for source \textit{content} tokens and a large-codebook tokenizer for \textit{content-style} tokens. VEVO mimics the style in an accent prompt and converts content into content-style tokens with the desired accent; a synthesizer then renders the waveform from the source style prompt. We use the official checkpoint.\footnote{\url{https://huggingface.co/amphion/Vevo}} The accent prompts are randomly drawn from the LibriTTS-R test set (\texttt{test-clean} and \texttt{test-other}), with utterances filtered to 4--12\,s and classifier-predicted US-accent probability $>99\%$.

\textbf{TokAN} (proposed): We report one resynthesis reference and two TokAN conversion modes (Sec.~\ref{sec:synthesizer}):
\begin{itemize}
\item \textit{Resynthesis}: Direct reconstruction (no conversion).
\item \textit{TokAN-1}: Duration-free mode after GRPO post-training.
\item \textit{TokAN-2}: Source-length mode after GRPO post-training.
\end{itemize}
For the token-conversion module we evaluate only the post-GRPO checkpoint in the main comparison; the effectiveness of post-training is demonstrated in the ablation study.

For FramAN, CosyAccent, and TokAN, we use the same pre-training and fine-tuning datasets (Sec.~\ref{sec:datasets}) while TokAN is additionally post-trained on unlabelled GLOBE.

\subsection{Evaluation Metrics}
\label{sec:metrics}

\textbf{Subjective metrics.}
We evaluate speech \textit{naturalness} (NAT) and \textit{accentedness} (ACT) using MUSHRA tests, and \textit{speaker similarity} (SIM) using best-worst scaling (BWS)~\cite{louviere2015best}.
BWS scores are aggregated via the counting algorithm~\cite{ravillion2020comparison}: $(N_\text{best} - N_\text{worst}) / N_\text{occurrence}$.
Each evaluation uses 15 test cases, where each case groups the source utterance with its converted counterparts from all systems for direct comparison; 25 raters score every case.
Native US samples are excluded from ACT, so ACT is computed on the six non-native L2-ARCTIC accents.

\textbf{Objective metrics.} Five objective metrics are deployed, one for intelligibility, one for speech quality, one for timbre preservation, and two for accentedness:
\begin{itemize}
\item \textit{Intelligibility / Content preservation}: Word error rate (WER, \%$\downarrow$) using the same native-only ASR model as mentioned in tokenizer selection (Sec.~\ref{sec:tokenizer_selection}).
\item \textit{Naturalness}: UTMOSv2 score ($\uparrow$)~\footnote{\url{https://github.com/sarulab-speech/UTMOSv2}} obtained from a neural MOS predictor~\cite{baba2024utmosv2}.
\item \textit{Timbre preservation}: Speaker encoding cosine similarity (SECS, $\uparrow$) using a pre-trained ECAPA-TDNN~\cite{desplanques2020ecapa} with a WavLM-Large backbone\footnote{\url{https://github.com/microsoft/UniSpeech/tree/main/downstreams/speaker_verification}}.
\item \textit{Accentedness reduction}: PPG distance ($\Delta$PPG, $\downarrow$) between generated and synthesized native target utterances~\cite{churchwell2024high}, measuring proximity to native pronunciation.
\item \textit{L1 probability} (L1-Prob, \%$\uparrow$): Sum of predicted confidence values over the five native English-accent labels of the classifier (US, England, Canada, Australia, New Zealand). Compared with the GRPO training reward---which sums only the US and England labels---this evaluation metric is more tolerant and better aligned with the overall target of L1 nativeness.
\end{itemize}

\subsection{Implementation Details}
\label{sec:implementation}

\textbf{Training hyperparameters.} The hyperparameters for the training recipe are summarized below.
\begin{itemize}
  \item \textit{Joint SSL tokenizer and synthesizer.} WavLM-Large layer-22 features; VQ codebook size 1024; $\beta=0.5$ in Eq.~\eqref{eq:vq_loss}; $\lambda_1 = 2.0$, $\lambda_2 = 2.0$, and $\lambda_3 = 0.5$ in Eq.~\eqref{eq:tok_loss}.
  \item \textit{BART pre-training.} Span masking probability $p_\text{mask} = 0.3$, Poisson span length $\lambda = 3.0$, random replacement probability within masked spans $p_\text{rand} = 0.3$, insertion probability $p_\text{ins} = 0.1$. Phonemic CTC loss weight $\gamma = 0.5$. Trained for 5 epochs on Emilia-EN. Optimizer: AdamW, learning rate 1e$-$4.
  \item \textit{SFT.} Trained for 3 epochs. Optimizer: AdamW, learning rate 5e$-$5.
  \item \textit{GRPO post-training.} Group size $G = 12$. Rollout sampling: top-$p$ = 0.85, top-$k$ = 25. Clip range $\epsilon = 0.2$.  KL penalty weight $\lambda_{\text{kl}} = 0.02$. Reward weights $w_{\mathrm{WER}} = 1.0$, $w_{\text{Acc}} = 0.5$. Trained for 40,000 update steps on GLOBE.
\end{itemize}

\textbf{Inference hyperparameters.} We deploy the strategy below for inference on the test set.
\begin{itemize}
  \item \textit{Token converter.} Beam size 10 for decoding.
  \item \textit{Synthesizer.} Euler sampler with 32 steps.  CFG strengths $w_1 = 1.0$, $w_2 = 1.0$ (Eq.~\eqref{eq:cfg}).  Duration predictor CFG strength $0.1$ (over the total-duration condition).
\end{itemize}

% ============================================================
%  X. RESULTS AND ANALYSIS
% ============================================================
\section{Results and Analysis}

\subsection{Tokenizer Analysis}

The tokenizer selection study (Sec.~\ref{sec:tokenizer_selection}) confirms that WavLM-Large layer-22 with codebook size 1024 achieves the best content preservation on the validation set.

The reconstruction WER---measuring how much a tokenizer's round-trip encoding reduces accentedness without any conversion---reveals an interesting property: tokenizers with stronger phonemic discriminability naturally cluster phonetically similar (accented and native) tokens together, providing an implicit initial normalization.

\subsection{Main Comparison}
\label{sec:main_comparison}

Table~\ref{tab:main_results} presents the main evaluation results.

\begin{table*}[t]
\centering
\caption{Evaluation results of accent normalization systems. Source-length ($\checkmark$) indicates whether the source total duration is preserved. Subjective scores are reported as mean $\pm$ 95\% confidence interval. Best and second-best objective results are in \textbf{bold} and \underline{underlined}, respectively.}
\vspace{-0.2cm}
\label{tab:main_results}
\small
\setlength{\tabcolsep}{5pt}
\begin{tabular}{lccccccccc}
\toprule
\multirow{2}{*}{System} & \multirow{2}{*}{Src-len} & \multicolumn{3}{c}{Subjective} & \multicolumn{5}{c}{Objective} \\
\cmidrule(r){3-5}\cmidrule(r){6-10}
& & NAT ($\uparrow$) & ACT ($\downarrow$) & SIM ($\uparrow$) & WER (\%$\downarrow$) & UTMOS ($\uparrow$) & SECS ($\uparrow$) & $\Delta$PPG ($\downarrow$) & L1-Prob (\%$\uparrow$) \\
\midrule
Source             & $\checkmark$ & 60.09{\tiny$\pm$2.38} & 47.39{\tiny$\pm$2.34} & --- & 15.81 & 3.04 & --- & 0.5092 & 74.06 \\
\midrule
FramAN~\cite{bai2024diffusion} & $\checkmark$ & 57.08{\tiny$\pm$2.36} & 43.89{\tiny$\pm$2.48} & -0.075 & 17.55 & 2.99 & 0.4478 & 0.4711 & 83.50 \\
CosyAccent-1~\cite{bai2026cosyaccent} & $\times$ & \underline{65.25}{\tiny$\pm$1.93} & 27.35{\tiny$\pm$1.84} & -0.075 & 12.40 & 3.22 & 0.3513 & 0.2734 & 90.04 \\
CosyAccent-2~\cite{bai2026cosyaccent} & $\checkmark$ & 58.87{\tiny$\pm$2.22} & 31.07{\tiny$\pm$2.05} & -0.096 & 13.84 & 3.12 & 0.3682 & 0.3027 & 87.24 \\
VEVO~\cite{zhang2025vevo} & $\checkmark$ & 62.03{\tiny$\pm$2.37} & 40.52{\tiny$\pm$2.55} & \underline{-0.023} & 28.94 & 3.01 & \underline{0.5775} & 0.5328 & 95.51 \\
\midrule
Resynthesis & $\checkmark$ & 60.54{\tiny$\pm$2.39} & 43.09{\tiny$\pm$2.29} & \phantom{-}\bf{0.417} & 14.01 & 3.20 & \bf{0.5862} & 0.4464 & 79.03 \\
TokAN-1         & $\times$ & {\bf70.73}{\tiny$\pm$1.95} & {\bf22.23}{\tiny$\pm$1.71} & -0.081 & \bf{9.23} & \bf{3.38} & 0.3655 & \bf{0.2533} & \bf{99.09} \\
TokAN-2         & $\checkmark$ & 62.90{\tiny$\pm$2.29} & \underline{25.51}{\tiny$\pm$1.99} & -0.067 & \underline{9.40} & \underline{3.26} & 0.3727 & \underline{0.2622} & \underline{99.01} \\
\bottomrule
\end{tabular}
\end{table*}

\textbf{Intelligibility / Content preservation.}
TokAN achieves substantially better intelligibility than all baselines, with TokAN-1 reaching a WER of 9.23\% and TokAN-2 at 9.40\%---both far below CosyAccent-1 (12.40\%) and FramAN (17.55\%).
This advantage is consistent across all accents (Table~\ref{tab:accent_wer}).
VEVO exhibits the highest WER (28.94\%), substantially worse than even the unconverted source (15.81\%).
We attribute this to its extremely small content tokenizer codebook (32 entries): while a small codebook may suffice for native speech, prior work has shown that at least 1000 codes are needed for robust content recognition~\cite{cui2025exploring1,yang2024towards,chang2024exploring}, and L2-accented speech---with its greater phonetic variability---exacerbates this bottleneck.
Although VEVO's second-stage acoustic tokenizer has 8192 entries, the initial content tokens already limit the information available for reconstruction.

\textbf{Accentedness reduction.}
TokAN achieves the strongest accent reduction across both objective and subjective metrics: the lowest $\Delta$PPG (0.2533 for TokAN-1), highest L1-Prob (99.09\%), and lowest ACT rating (22.23).
An interesting discrepancy arises with VEVO: it achieves a high L1-Prob (95.51\%) despite having the worst $\Delta$PPG (0.5328) and ACT (40.52).
This is because L1-Prob only measures whether the surface pronunciation patterns resemble native accents, regardless of whether they correspond to the intended content.
This highlights the importance of jointly evaluating accent reduction with content fidelity.

\textbf{Naturalness.}
TokAN-1 achieves the highest naturalness (NAT = 70.73), outperforming all baselines.
TokAN-2 receives a notably lower naturalness score (62.90) despite comparable objective quality (UTMOS = 3.26 vs.\ 3.38).
This is likely because the source-length constraint forces certain syllables to be prolonged to match the L2 speaker's slower rhythm, and raters perceive such prolongation as unnatural even when the overall audio quality is high---an observation consistent with our preliminary findings~\cite{bai2025accent}.

\textbf{Speaker similarity.}
Resynthesis achieves the highest speaker similarity (SIM = 0.417, SECS = 0.5862), as it performs no phonetic conversion and thus preserves maximal speaker characteristics.
Among accent normalization systems, VEVO achieves the second-best similarity (SIM = $-$0.023, SECS = 0.5775), followed by TokAN-2 (SIM = $-$0.067, SECS = 0.3727).
% and TokAN-1 (SIM = $-$0.081, SECS = 0.3655).
Comparing the three TokAN conditions (Resynthesis, TokAN-2, TokAN-1) reveals a general trade-off between speaker similarity and accent reduction: as $\Delta$PPG decreases (0.4464 $\to$ 0.2622 $\to$ 0.2533), speaker similarity also decreases. This observation is consistent with findings in \cite{huang2026codecmos}, suggesting that some speaker-identifying characteristics could be entangled with accent-related features at the token level.
VEVO's relatively high speaker similarity likely stems from two factors: (i)~it preserves the source rhythm without modification, and (ii)~it employs a prompt-based speech synthesizer with stronger voice-cloning ability.
This suggests that replacing the current embedding-conditioned synthesizer with a prompt-based one may achieve a better balance between accentedness reduction and speaker identity preservation, which we leave for future work.

\subsection{Accent-wise Analysis}

\begin{table}[!t]
\centering
\caption{Accent-wise WER (\%) with native-only ASR. Best results per accent in \textbf{bold}.}
\label{tab:accent_wer}
\small
\setlength{\tabcolsep}{3pt}
\begin{tabular}{lccccccc}
\toprule
System & Ar & Zh & Hi & Ko & Es & Vi & Us \\
\midrule
Source       & 15.02 & 20.76 & 11.63 & 13.69 & 15.95 & 31.11 & \underline{2.50} \\
\midrule
FramAN       & 18.92 & 21.74 & 15.57 & 14.81 & 27.61 & 28.61 & 5.62 \\
CosyAccent-1 & 13.76 & 16.26 & 8.3 & 10.41 & 12.98 & 21.35 & 3.15 \\
CosyAccent-2 & 14.98 & 18.33 & 9.62 & 12.09 & 15.14 & 23.55 & 3.22 \\
VEVO & 28.92 & 36.65 & 33.27 & 23.58 & 33.76 & 41.95 & 5.95 \\
\midrule
Resynthesis  & 14.00 & 18.03 & 9.52 & 11.76 & 14.37 & 28.48 & \bf{1.94} \\
TokAN-1   & \bf{9.39} & \bf{11.89} & \bf{5.85} & \bf{7.65} & \bf{8.90} & \bf{18.10} & 2.83 \\
TokAN-2   & \underline{9.76} & \underline{12.02} & \underline{5.48} & \underline{7.78} & \underline{9.11} & \underline{18.63} & 2.99 \\
\bottomrule
\end{tabular}
\end{table}

Table~\ref{tab:accent_wer} reports per-accent WERs using the native-only ASR model.
TokAN achieves the best WER for all six L2 accents. Notable patterns include:
\begin{itemize}
\item \textit{Chinese and Vietnamese}: Large improvements in WER, attributed to TokAN's ability to normalize syllable-timed rhythmic patterns from these L1 languages, which have a strong influence on L2 English timing. A detailed phonemic analysis for Chinese is provided in Sec.~\ref{sec:phoneme_analysis}.
\item \textit{Native US English}: Near source-level WER, confirming that the system correctly preserves native speech without degradation when no accent conversion is needed.
\end{itemize}

\subsection{Duration Control Analysis}

\begin{figure}[!t]
\centering
\includegraphics[width=0.95\columnwidth]{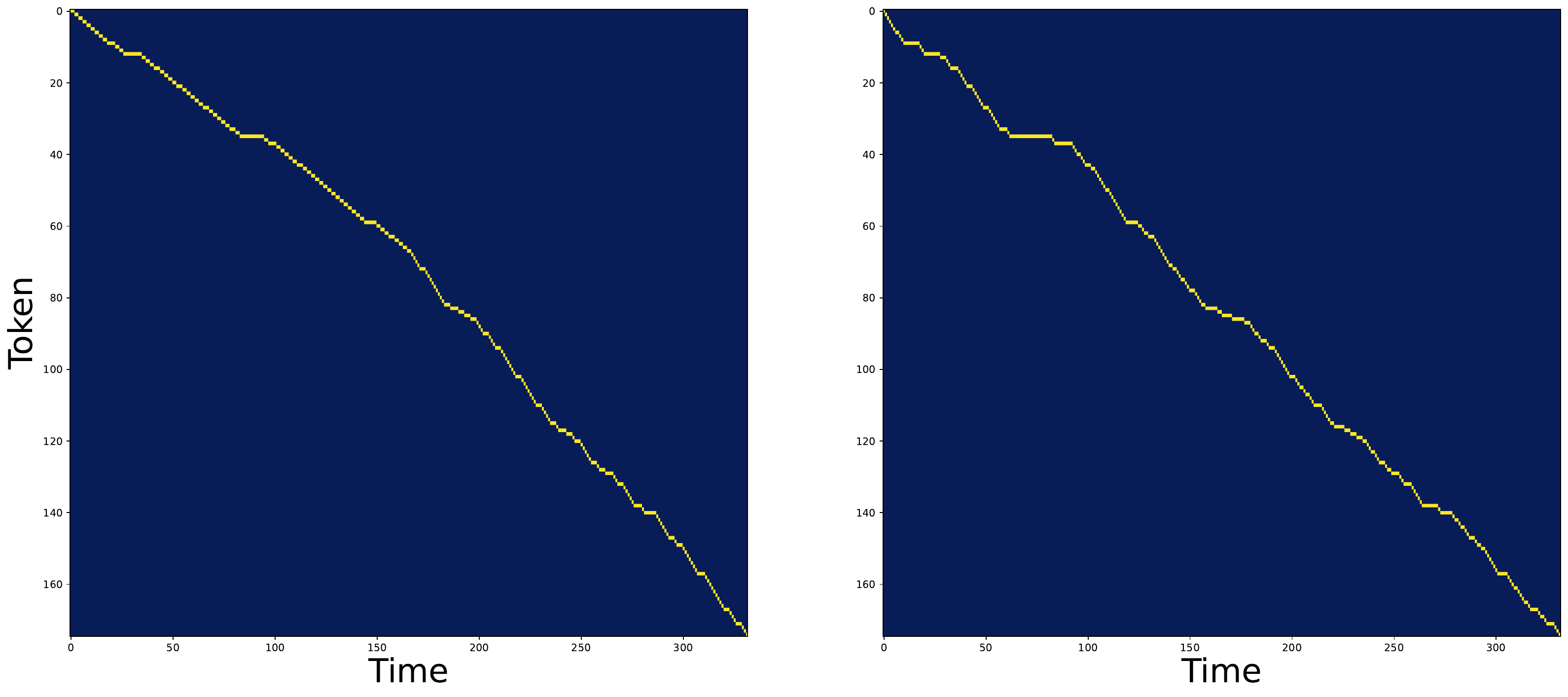}
\caption{Token-frame alignment comparison. Left: TokAN-1 with direct duration scaling to match source total duration. Right: TokAN-2 with total-duration-aware prediction. The scaling approach produces uniform durations (diagonal alignment), while the duration predictor allocates varied durations across tokens, yielding more natural prosody.}
\label{fig:duration}
\end{figure}

Figure~\ref{fig:duration} illustrates token-frame alignments produced by the two duration control strategies.
Direct scaling (TokAN-1 with source-length matching) produces more uniform, smooth durations, while the total-duration-aware predictor (TokAN-2) yields more varied token durations---particularly for vowels and fricatives---resulting in more natural prosody.

\begin{table}[!t]
\caption{Duration control analysis. ``Src-len'' indicates whether source total duration is enforced. Arrows ($\rightarrow$) mark the default operating mode of each system. Duration difference is measured against source utterance duration.}
\label{tab:duration}
\centering
\small
\setlength{\tabcolsep}{5pt}
\begin{tabular}{llccc}
\toprule
\multirow{2}{*}{System} & \multirow{2}{*}{\phantom{ }Src-len} & \multicolumn{2}{c}{Duration Difference} & \multirow{2}{*}{$\Delta$PPG ($\downarrow$)} \\
\cmidrule(r){3-4}
& & Abs. (s$\downarrow$) & Rel. (\%$\downarrow$) \\
\midrule
\multirow{2}{*}{TokAN-1} & $\rightarrow$ $\times$ & 0.79 & 19.18 & 0.2533 \\
                              & \phantom{$\rightarrow$} \checkmark & - & - & 0.2705 \\
\midrule
\multirow{2}{*}{TokAN-2} & \phantom{$\rightarrow$} $\times$ & 0.07 & 1.64 & 0.2608 \\
                              & $\rightarrow$ \checkmark & - & - & 0.2622 \\
\bottomrule
\end{tabular}
\end{table}

Table~\ref{tab:duration} quantifies the effect of duration control on both total-duration fidelity and accent reduction.
TokAN-1 operates in duration-free mode by default: it predicts native-like token durations without any total-duration constraint, resulting in a large deviation from source durations (0.79\,s absolute, 19.18\% relative).
When direct scaling is applied to force source total duration, TokAN-1's $\Delta$PPG increases from 0.2533 to 0.2705, indicating that uniform scaling distorts the natural prosodic structure learned by the model.

In contrast, TokAN-2 operates with total-duration-aware prediction by default.
Even without a final scaling step, its predicted durations already closely match the source total duration (only 0.07\,s absolute, 1.64\% relative deviation), thanks to the duration predictor being conditioned on the source average token duration.
Applying final scaling to TokAN-2 causes only a negligible change in $\Delta$PPG (0.2608 $\to$ 0.2622), confirming that the predictor has already distributed durations in a manner consistent with the source total length.
Crucially, TokAN-2 with source-length matching achieves a lower $\Delta$PPG (0.2622) than directly scaled TokAN-1 (0.2705), demonstrating that the total-duration-aware predictor generates more natural prosody while maintaining the source total duration---a desirable property for dubbing applications where duration fidelity is required.

\subsection{Ablation Study}

Table~\ref{tab:ablation} presents ablation results on the conversion model, using TokAN-1 (duration-free mode) as the base configuration.

\begin{table}[!t]
\centering
\caption{Ablation study. All systems use TokAN-1 (duration-free mode). Best results in \textbf{bold}.}
\label{tab:ablation}
\small
\setlength{\tabcolsep}{4pt}
\begin{tabular}{lccc}
\toprule
System & WER (\%$\downarrow$) & SECS ($\uparrow$) & $\Delta$PPG ($\downarrow$) \\
\midrule
Source              & 15.81 & --- & 0.5092 \\
\midrule
TokAN-1          & \bf{9.23} & 0.3655 & \textbf{0.2533} \\
\quad w/o post-training & 9.89 & 0.3791 & 0.2589 \\
\qquad w/o fine-tuning & 14.92 & \textbf{0.5605} & 0.4477 \\
\qquad w/o pre-training & 12.67 & 0.3894 & 0.2887 \\
\quad\qquad w/o CTC sup. & 12.91 & 0.3846 & 0.2907 \\
\bottomrule
\end{tabular}
\end{table}

\textbf{Effect of RL post-training.}
Removing the GRPO stage (``w/o post-training'') increases WER from 9.23\% to 9.89\% and $\Delta$PPG from 0.2533 to 0.2589.
This confirms that task-level RL, directly optimized intelligibility and accent reduction, further improves the accent normalization performance.
Meanwhile, SECS slightly increases without RL (0.3655 $\to$ 0.3791), suggesting that RL's accent-focused optimization mildly trades off speaker similarity.

\textbf{Effect of SFT (pre-training only).}
Removing the SFT stage (``w/o fine-tuning'') and relying solely on BART-style pre-training leads to substantial degradation: it works similar to direct resynthesis while being slightly worse.
This confirms that BART pre-training alone is insufficient for accent conversion---fine-tuning on parallel token pairs is essential for learning the L2-to-L1 phonetic mapping.

\textbf{Effect of pre-training.}
Removing the BART pre-training stage (``w/o pre-training'') while retaining SFT increases WER from 9.89\% to 12.67\% and $\Delta$PPG from 0.2589 to 0.2887.
This demonstrates that BART pre-training provides crucial language modeling priors over token distributions, enabling the model to better generalize during fine-tuning.

\textbf{Effect of CTC guidance.}
Further removing the CTC phoneme supervision from the encoder (``w/o CTC sup.'') increases WER to 12.91\%, confirming the finding from~\cite{bai2025accent}: explicit phonemic guidance is helpful for maintaining content fidelity during accent conversion.

\subsection{Token-Level Phonemic Distribution Analysis}
\label{sec:phoneme_analysis}

To evaluate pronunciation normalization at the token level, we analyze phoneme-specific token distributions using forced alignment~\cite{mcauliffe2017montreal} on the test set.
We compute the KL divergence between the per-phoneme token distributions of converted speech and those of synthetic native targets.

\begin{figure*}[t]
\centering
\includegraphics[width=0.8\textwidth]{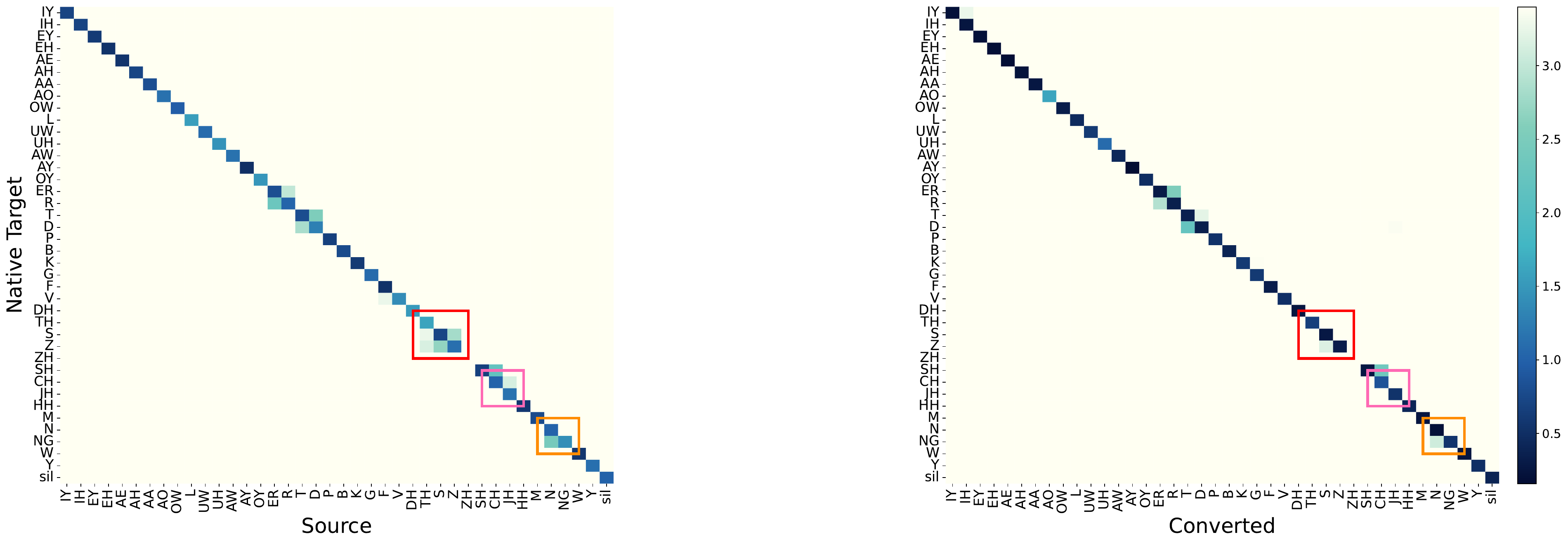}
\caption{Token-based phoneme divergence from native US English targets. Each cell shows the KL divergence between the token distribution of a given phoneme (row) in the analyzed speech and the native reference. Left: source Chinese-accented speech. Right: converted speech. After conversion, the diagonal becomes notably darker (lower KLD), indicating closer alignment with native pronunciations. Colored annotations highlight three prominent Mandarin-influenced mispronunciation patterns: dental-alveolar fricative confusion among /TH/, /S/, and /Z/ (red), devoicing of /JH/ to /CH/ (pink), and 
confusion between /N/ and /NG/ due to nasal coda weakening (orange).}
\label{fig:phoneme_dist}
\end{figure*}

Figure~\ref{fig:phoneme_dist} presents this analysis for Chinese-accented speech, before and after conversion with TokAN-1.

The post-conversion results show a significantly darker (lower-valued) diagonal, with the average on-diagonal KL divergence reduced from 1.076 to 0.519 (a 51.7\% reduction), indicating successful transformation toward native-like pronunciations.
Three prominent Mandarin-influenced English pronunciation patterns are highlighted:
\begin{itemize}
\item \textit{Dental-alveolar fricative confusion} (red): Mandarin lacks the English dental fricatives /TH/ and /DH/, causing speakers to substitute /S/ or /Z/. This manifests as high off-diagonal KLD among /TH/, /S/, and /Z/. After conversion, TokAN restores the dental fricative distinctions.
\item \textit{Affricate devoicing} (pink): Mandarin does not contrast voiced and voiceless affricates, leading speakers to realize /JH/ (as in ``judge'') as its voiceless counterpart /CH/ (as in ``church''). Post-conversion token distributions show recovered /JH/--/CH/ separation.
\item \textit{Nasal coda weakening} (orange): Mandarin contrasts pinyin \textit{-n} and \textit{-ng}, but final nasal codas can be realized with incomplete oral closure, leaving vowel quality and nasalization as major perceptual cues~\cite{duanmu2007phonology}. This can make English /N/ and /NG/ perceptually closer in Chinese-accented speech, which appears as elevated off-diagonal KLD between the two nasal categories. TokAN reduces this overlap and shifts the /N/ distribution toward the native alveolar nasal pattern.
\end{itemize}
These results demonstrate that TokAN operates at a phonetically meaningful level, correcting systematic L1-transfer patterns rather than merely applying surface-level transformations.
We note that the relative KLD reduction is less pronounced than in the preliminary conference version~\cite{bai2025accent}, likely because \textit{the improved jointly trained VQ tokenizer already performs a degree of implicit accent normalization during tokenization}---the source tokens are closer to native targets to begin with, which is consistent with the overall performance improvements observed in the main evaluation.

% ============================================================
%  XI. CONCLUSION
% ============================================================
\section{Conclusion}

We presented \textbf{TokAN}, a token-based accent normalization framework that operates on self-supervised discrete speech tokens to convert L2-accented speech into native-like speech while preserving speaker identity.
TokAN avoids the need for real parallel L1--L2 speech by using semi-synthetic token pairs for supervised fine-tuning, and further adapts using unpaired real multi-accented speech through GRPO post-training.
Building on the preliminary conference version~\cite{bai2025accent}, this paper introduces three main advances: (i)~a jointly trained VQ tokenizer that optimizes codebook assignments for speech synthesis and phonetic content simultaneously; (ii)~an updated autoregressive encoder-decoder with RoPE and a self-attention-only decoder, removing accent embedding in favor of an accent-universal architecture; and (iii)~a GRPO-based RL post-training stage that directly optimizes WER and accent classifier rewards without requiring additional paired data.
Experiments on seven English accents demonstrate that TokAN achieves the best WER of 9.23\% and a native accent probability of 99.09\%, outperforming a frame-to-frame baseline (FramAN), a direct flow-matching baseline (CosyAccent), and a prompt-based token conversion baseline (VEVO) across the key metrics of content preservation and accentedness reduction.
A flow-matching duration predictor with total-duration conditioning further supports dubbing and other duration-sensitive applications.

Several directions remain for future work.
First, our analysis in Sec.~\ref{sec:main_comparison} suggests that the embedding-conditioned synthesizer is a bottleneck for speaker similarity; replacing it with a prompt-based synthesizer that conditions on a reference utterance from the source speaker may achieve a better balance between accentedness reduction and speaker identity preservation.
Second, extending RL post-training with speaker similarity rewards may further improve timbre preservation without paired data.
Finally, the modular architecture of TokAN could be extended to non-English accent normalization by swapping the tokenizer and synthesizer components.

% ============================================================
%  ACKNOWLEDGMENT
% ============================================================
\section*{Acknowledgment}
This research is supported by National Natural Science Foundation of China (Grant No. 62401377 and No. 62271432), Program for Guangdong Introducing Innovative and Entrepreneurial Teams (Grant No. 2023ZT10X044), Yangtze River Delta Science and Technology Innovation Community Joint Research Project (Grant No. 2024CSJGG1100), Shenzhen Science and Technology Program (Shenzhen Key Laboratory, Grant No. ZDSYS20230626091302006), Shenzhen Stability Science Program 2023, Shenzhen Key Lab of Multi-Modal Cognitive Computing, and the internal project of the Guangdong Provincial Key Laboratory of Big Data Computing (Grant No. B10120210117-KP02), The Chinese University of Hong Kong, Shenzhen (CUHK-Shenzhen).

% ============================================================
%  REFERENCES
% ============================================================
\bibliographystyle{IEEEtran}
\bibliography{refs}

\end{document}